# Energy-Adaptive Checkpoint-Free Intermittent Inference for Low Power Energy Harvesting Systems

Sahidul Islam*, Wei Wei *, Jishnu Banerjee*, Chen Pan†
*Department of Computer Science, The University of Texas at San Antonio
†Department of Electrical and Computer Engineering, The University of Texas at San Antonio

*Abstract*—Deep neural network (DNN) inference in energy harvesting (EH) devices poses significant challenges due to resource constraints and frequent power interruptions. These power losses not only increase end-to-end latency, but also compromise inference consistency and accuracy, as existing checkpointing and restore mechanisms are prone to errors. Consequently, the quality of service (QoS) for DNN inference on EH devices is severely impacted. In this paper, we propose an energy-adaptive DNN inference mechanism capable of dynamically transitioning the model into a low-power mode by reducing computational complexity when harvested energy is limited. This approach ensures that end-to-end latency requirements are met. Additionally, to address the limitations of error-prone checkpoint-and-restore mechanisms, we introduce a checkpoint-free intermittent inference framework that ensures consistent, progress-preserving DNN inference during power failures in energy-harvesting systems.

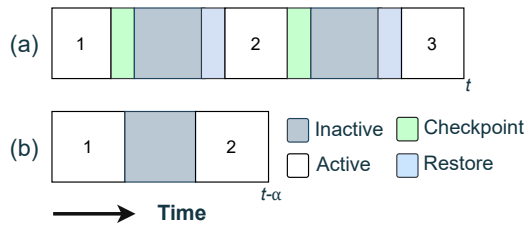

Fig. 1: Inference latency in low-power mode. (a) Existing methods involve additional power cycles and checkpoint/restore overhead. (b) Our proposed checkpoint-free intermittent inference adapts to low-power mode, reducing power cycles and end-to-end latency while gurantee consistency.

## I. INTRODUCTION

Energy Harvesting (EH) technology is a promising alternative that harnesses power from the environment, reducing carbon emissions by minimizing reliance on traditional battery technology [1]. With the abundance of renewable energy sources like solar, thermal, and wind, EH presents a sustainable and environmentally friendly solution compared to current battery-based systems [2]–[4]. These energy sources are readily available and enviromnet friendly, making them ideal for powering IoT and embedded devices [5]. EH-powered devices require less maintenance, are cost-effective, and offer long-term reliability, making them suitable for a wide range of applications, including smart cities, smart agriculture, monitoring systems in wildlife or remote locations.

However, to build such an effective system with IoT and embedded devices, it is not enough for these devices to simply collect and transmit data. It is equally important for them to process the data locally and make intelligent decisions in real time [5]–[7]. This level of intelligence enables IoT systems to adapt to dynamic environments, optimize operations, and provide actionable insights without relying solely on cloud services. Achieving such intelligent decision-making requires the deployment of highly efficient algorithms capable of running directly on resource-constrained devices. Among the various artificial intelligence techniques, Deep Neural Networks (DNNs) have emerged as one of the most prominent and successful approaches, revolutionizing fields such as image recognition, natural language processing, and predictive analytics.

Nevertheless, deploying Deep Neural Network (DNN) algorithms on energy-harvesting-powered devices presents significant challenges due to the inherent limitations of such systems [8]–[10]. The problems are multifaceted and require innovative solutions to address them effectively.

**Challenge 1:** Firstly, energy-harvesting-powered devices are typically extremely resource-constrained, featuring limited hardware capabilities. These devices often have only kilobytes of memory, low processing power with minimal CPU frequency, and lack advanced computational resources. On the other hand, DNN algorithms are inherently resource and computation-intensive, requiring significant memory for storing parameters and intermediate computations, as well as substantial processing power for executing operations like matrix multiplications, convolutions, and activation functions. The disparity between the computational demands of DNNs and the hardware constraints of energy-harvesting devices makes it a grand challenge to deploy DNNs in such systems without compromising the quality of service (QoS). For instance, executing large DNN models in constrained environments may lead to delays, reduced accuracy, or complete failures in time-sensitive applications.

As illustrated in Figure 1 (a), under conditions of low harvesting power, completing an inference task may require multiple power cycles, potentially failing to meet Quality of Service (QoS) requirements. To ensure QoS is maintained, it is crucial to achieve energy adaptivity by reducing computational complexity when power levels are low. A straightforward solution would be to activate a smaller model when low harvesting situation occurs. However, for energy-harvesting (EH) devices with limited resource budgets, it is impractical to store and activate multiple models of varying sizes in real time. This limitation makes such an approach unfeasible. To tackle this challenge, we propose dynamically adaptive DNN inference approach. By leveraging pattern-based pruning and shared weight design method, the system can seamlessly transition to a power-saving mode for inference without the need to deploy additional smaller models on the device. This allows the system to efficiently adapt to varying power conditions while maintaining performance.

**Challenge 2:** Secondly, energy-harvesting devices rely on intermittent and unpredictable energy sources, such as solar or vibration energy, making them prone to frequent power interruptions. These interruptions can halt computations mid-process, leading to loss of progress and potentially requiring the system to restart computations from scratch once power is restored. This is particularly problematic for DNN algorithms, as they involve sequential layers of computations that cannot simply resume without losing integrity. Restarting the entire computation repeatedly not only wastes energy but also significantly affects the efficiency and reliability of the system. Hence, it is crucial to design solutions that enable the system to gracefully handle power interruptions by resuming computations from the point of failure without losing progress.

Moreover, the lack of sophisticated operating systems in such devices adds another layer of complexity. Energy-harvesting-powered devices often operate on lightweight or custom firmware rather than full-fledged operating systems. This means the responsibility of handling power failures, resource allocation, and computational efficiency falls largely on the deployed algorithm itself. Therefore, any DNN algorithm designed for these systems must be aware of and resilient to power failure situations [11], [12].

To tolerate power failures in energy-harvesting-powered IoT systems, existing checkpoint-based methods [13], [14], while widely used, present significant challenges when applied to Deep Neural Network (DNN) algorithms. These methods, introduce substantial overhead, resulting in considerable delays in the end-to-end latency of DNN computations. This overhead is primarily caused by the frequent need to save and restore computation states, especially for large and complex DNN models, which involve extensive parameters and activations. Some other check-pointing mechanisms heavily rely on the accurate prediction of available energy to determine when to save computation states. The inability to predict power failures effectively can lead to substantial progress loss and computation inconsistencies, ultimately increasing latency and resulting in inaccurate inference.

We propose that traditional checkpointing and restore mechanisms may not be the most efficient approach for DNN algorithms in energy-harvesting scenarios. Instead, a more efficient checkpointless method could be designed to address these challenges. By leveraging the structural properties of DNN computations, which are largely composed of multi-level loops iterating over layers, weights, and activations, we can redefine how progress is tracked and computation is resumed. In this approach, the index of the loops within the DNN computation can act as a sufficient representation of progress, allowing the system to resume computations seamlessly after a power failure without the need for heavy checkpointing.

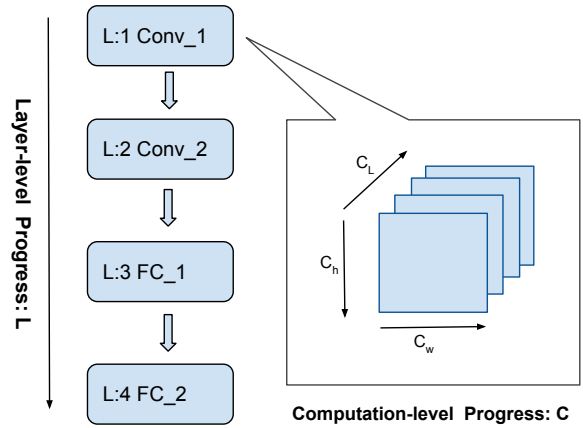

Fig. 2: Checkpoint-Free Intermittent Inference Via Multi-level Progress Preservation.

This checkpoint-free method is feasible because most DNN weights and activations are already stored in non-volatile memory (NVM) in energy-harvesting-powered devices. NVM ensures that the essential data required for computation, such as model weights and intermediate results, remains intact even during power outages. By storing only minimal metadata, such as loop indices, the system can resume computation from the exact point of interruption. This eliminates the need to save and restore large amounts of data, significantly reducing overhead and improving end-to-end latency.

However, the loop index-based checkpoint-free approach also comes with its own set of challenges. We discovered that it can potentially introduce the write-after-read (WAR) problem. This issue arises when a multiply-accumulate (MAC) operation within a convolution or fully connected layer fails to be followed by the corresponding set of increment operations due to an unexpected power failure. In such a scenario, the incomplete execution of these operations can disrupt the computational flow, leading to inconsistent computations during inference. As a result, the system may produce incorrect or unreliable results, compromising the accuracy and reliability of the inference process. To address this, we proposed a simple yet effective solution that leverages energy prediction to mitigate the problem.

The major contributions are summarized as follows.

- We propose a checkpoint-free intermittent computation paradigm that leverages the loop index in DNN computations stored in non-volatile memory. This approach eliminates the need for frequent checkpointing, which is typically required in traditional intermittent computation systems.

- We further investigated the proposed method and identified the write-after-read (WAR) problem as a an inherent to its execution. To address this issue, we isolated the critical regions of the computation and ensured their execution as a cohesive unit by leveraging energy prediction.
- We propose low energy adaptivity (LEA) mechanism that enables the system to dynamically adjust its computational workload in response to low harvesting power, eliminating the need for additional model overhead

The rest of this paper is organized as follows. Section 2 describes the proposed framework. Section 3 states the experimental evaluation and results. Lastly, Section 4 provides a conclusion.

## II. METHODOLOGY

### A. Checkpoint-Free Intermittent Inference

In intermittent systems, it is crucial to preserve computational progress due to frequent power failures. When power is interrupted, the system must be able to resume computations precisely from where it left off, ensuring no loss of progress once power is restored. In the context of convolutional neural network (CNN) inference, progress is accumulated through a series of multiplication and addition operations executed across multiple nested loops. We propose that, rather than relying on frequent checkpointing, the progress of CNN inference can be effectively tracked using the index of the convolution loop and the current layer being processed. As illustrated in Figure 2, progress preservation is divided into two components: 1) Layer-level progress, which tracks the active layer in the network, and 2) Computation-level progress, which captures the precise state of the loop index within the convolution operation.

*1) Layer-level progress:* Layer-level progress is denoted by $L$, stores the rank or ID of the currently executing layer. Once the execution of a layer is completed, the progress variable is updated with the rank or ID of the subsequent layer to be executed. This approach eliminates the need for additional overhead, such as frequent checkpointing. In the event of a power failure, the system simply reads the rank or ID of the currently executing layer from non-volatile memory (NVM) upon power restoration and resumes computation seamlessly.

*2) Computation-level progress:* Computation-level progress is denoted by $C$, tracks the index of the input, weight, and output memory locations during the computation process. At the start of the computation, the system retrieves the input data and weights from non-volatile memory (NVM). As the computation progresses, intermediate results are generated, and the final output is written back to NVM. The computation-level progress, $C$, is defined by the triple $(I,W,O)$, which represents the current execution indices of the input, weight, and output data, respectively. This triple plays a crucial role in pinpointing the exact location where the computation was interrupted during a power failure.

When power is restored, the system retrieves the stored computation-level progress triple $C(I,W,O)$ from NVM to identify the last computed memory location. This enables the system to resume computation precisely from where it was interrupted, ensuring that no data is lost or unnecessarily recomputed.

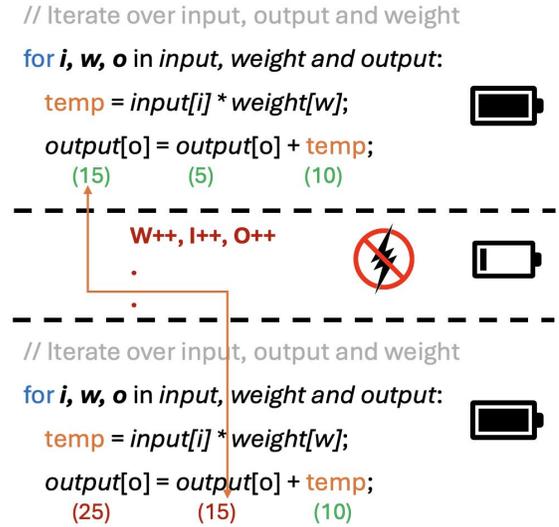

Fig. 3: Example of a WAR hazard in intermittent computation. Failing to change loop-indices due to power failure causing data inconsistency when power resumes.

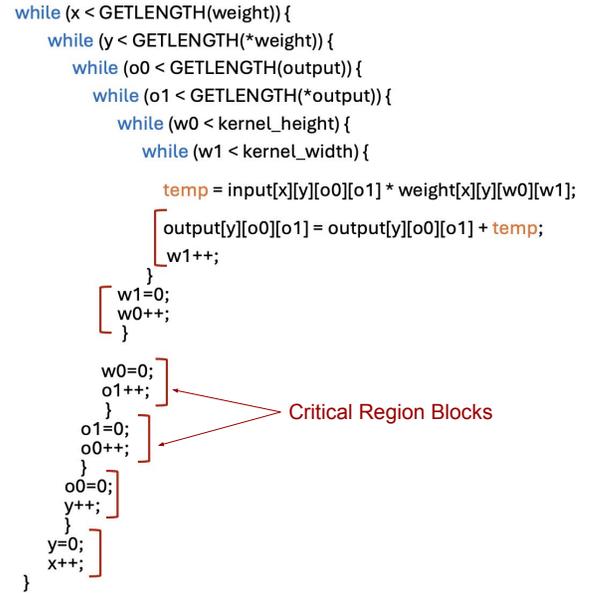

Fig. 4: Several Critical Section Blocks in Convolutional Layer

Therefore, unlike traditional checkpointing or tile-based mechanisms, this approach eliminates the need to store additional information such as intermediate results or predict potential power failure scenarios. Furthermore, unlike tile-based mechanisms, it does not require partitioning of the CNN, making it significantly more convenient and straightforward to implement.

### B. Consistency-aware idempotent execution

Although the index-based checkpoint-free progress preservation offers an efficient approach for intermittent inference,

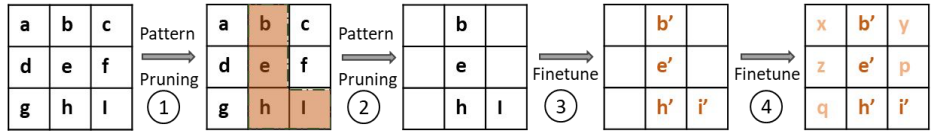

Fig. 5: Weight concentration during training is achieved through a two-step fine-tuning process. First, kernel pattern pruning is applied, followed by fine-tuning. In the next step, only the pruned weights undergo further fine-tuning.

our investigation revealed that the system can still fall into an inconsistent state, leading to inaccurate results. Specifically, we identified the potential occurrence of a write-after-read (WAR) hazard, as illustrated in Figure 3. Previous approaches [6], such as the double buffering mechanism, attempted to address this issue by requiring multiple data swaps before finalizing intermediate results. In this method, filter data is first swapped to temporary memory for computation, and the results are later swapped back to their required locations. However, due to the high data movement cost, this solution significantly increases inference latency and proves to be inefficient for energy-constrained systems.

To overcome this challenge, we propose a Consistency-aware Idempotent Execution strategy. First, we analyze and identify all critical section blocks within the convolution code that are susceptible to write-after-read (WAR) hazards and can lead to data inconsistency during power failures. These critical regions are tightly coupled, meaning that partial execution due to a power failure can result in WAR hazards. By isolating these critical sections and ensuring their uninterrupted execution, we guarantee that the inference process becomes idempotent, meaning it can be repeated multiple times across power cycles without altering the final result.

*1) WAR in DNN inference:* A DNN inference is susceptible to write-after-read (WAR) hazards when a power failure occurs mid-computation. As illustrated in Figure 3, data is both read from and written to the same output location. If a power failure happens before incrementing the pointer variable, the system will resume computation in the next power cycle using the stored layer-level progress L and computation-level progress C. However, because the pointer update was interrupted, the system reads the modified data instead of the original data, leading to data inconsistency. In this example of figure 3, the output data is expected to be 5 after power resumes. However, due to the power failure, the computation-level progress $C(I,W,O)$ fails to update the pointer. As a result, when the computation resumes, the output is recalculated using the previously written result (15), leading to inconsistent and incorrect data.

*2) Critical Sections and Tightly-coupled Instruction set:* We identified several critical sections in a convolution, as shown in Figure 4. The instructions within these sections are tightly coupled instruction set, meaning that within the same power cycle, executing one instruction necessitates executing the others. Otherwise, the inference results will become non-idempotent, causing the system to encounter a Write-After-Read (WAR) scenario as discussed before.

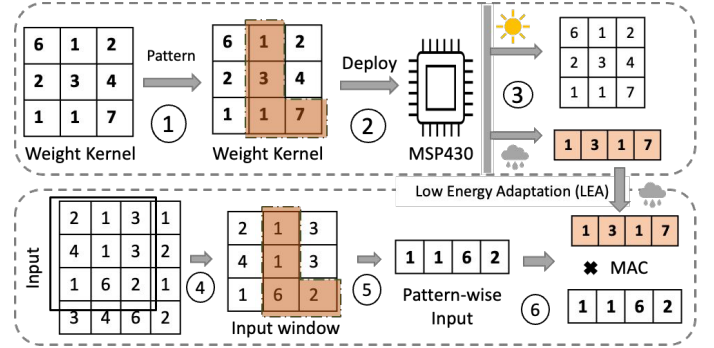

Fig. 6: Low Energy Adaption method. Pattern-wise weights are concentrated and deployed. Under low energy conditions, only the concentrated weights are activated to perform computations, ensuring energy-efficient inference

To prevent WAR scenario, the tightly coupled instruction sets within critical sections must execute atomically within the same power cycle. To ensure this, our approach allows the program to enter a critical section only when sufficient energy is available for its execution. Specifically, if the energy required to execute the instructions within a critical section is $E_{req}$ and the available energy of the device is $E_{av}$, then execution is permitted only if $E_{av} > E_{req}$.

### C. Low Energy Adaptation (LEA)

To adapt with low-energy conditions, we introduce an adaptation mechanism that leverages a pattern-based weight concentration technique, eliminating the need for additional model actuation on the energy-harvesting (EH) device. The process is divided into two phases: training and inference.

*1) Training:* In addition to standard DNN training, we implement a two-step fine-tuning process to concentrate important weights within a smaller region of each kernel or block, dictated by a predefined pattern. This pattern is selected based on L1 normalization [15], [16]. The key idea is that only the pattern-concentrated weights are activated in low-power mode, removing the need for an additional model. As illustrated in Figure 7, we first apply pattern-based pruning at the kernel level, followed by fine-tuning to restore accuracy. Next, only the pruned elements undergo further fine-tuning and the pattern-based concentrated weights which are already fine-tuned are kept frozen so that they can not get changed during second level fine-tuning. The idea is that only the concentrated weights will be acuated during low harvesting scenario.

*2) Inference:* During inference, the trained model with concentrated weight patterns is deployed on the device. When

TABLE I: Convolutional Neural Network Dataset and Model Detail

| Tasks | Layer | Model Architechture | Concentraton Method | Concentration Ratio | Concentrated Size |
|---|---|---|---|---|---|
| Image Classification (MNIST) | Conv | 8 x 1 x 5 x 5 | Pattern | 58.8% | 19k |
| | Conv | 16 x 8 x 3 x 3 | Pattern | 33.3% | |
| | FC | 100 x 400 | Pattern | 44.44% | |
| | FC | 10 x 100 | — | — | |
| Human Activity Recognition (HAR) | Conv | 8 x 1 x 1 x12 | — | — | 14k |
| | FC | 64 x 880 | Pattern | 24.3% | |
| | FC | 6 x 64 | — | — | |
| German Traffic Sign Recognition Benchmark (GTSRB) | Conv | 16x3x3x3 | — | — | 33k |
| | Conv | 16x4x3x3 | Pattern | 44.44% | |
| | FC | 100x512 | Pattern | 55.55% | |
| | FC | 43x100 | — | — | |

energy availability is sufficient, the full model is utilized for standard computation. However, under low-energy conditions, only the pattern-concentrated weights are activated, participating in the convolution process, as shown in Figure 6. Similarly, in the fully connected layer, the large weight matrix is divided into several smaller blocks, each treated as a kernel in the convolutional layer for low-energy adaptation and computation.

### III. EXPERIMENTAL EVALUATION

In this section, we will evaluate the performance of the proposed DNN architecture in terms of accuracy, latency and memory footprint.

#### A. Experimental Setup

**Hardware Setup:** We implemented our models on TI's MSP430FR5994 ultra-low-power evaluation board, which consists of a 16 MHz MCU, a 8KB volatile SRAM, a 256KB nonvolatile FRAM memory, and a low-energy Accelerator (LEA) that's operation independent of CPU for Signal Processing. The FRAM technology combines the low-energy fast writes, flexibility, and endurance of SRAM with the nonvolatile behavior of flash. In SRAM, 4KB is shared between CPU and LEA. The LEA accelerator can efficiently process data using complex functions including FFT, IFFT, and MAC. Energy is buffered with a capacitor of 100μF. For the energy measurement, we used CCS energy trace technology [17].

**Dataset and Model**: This paper considers three Different Dataset for Image Classification (MNIST) [18], Human Activity Recognition (HAR) [19], and German Traffic Sign Recongnition (GTSRB) [20] which represent image-based applications and wearable applications, as shown in Table I.

#### B. Experimental Results

We evaluated our framework by comparing it with the BASE model and the M1 model. The BASE model is defined as a standard model that lacks both Low Energy Adaptivity (LEA) and Checkpoint-Free Intermittent (CLI) execution methods, making it less efficient under energy-constrained conditions. The M1 model serves as another baseline for comparison, specifically to assess memory consumption, representing the memory requirements in a traditional setup.

*1) DNN model training and architecture detail:* We first train a custom model for each dataset since TI's MSP430FR5994 has only 256 KB of FRAM, making it impractical to deploy established models on such a resource-constrained device. The details of the individual models for each dataset are shown in Table I.

The model architecture for the MNIST and GTSRB datasets consists of two convolutional (CONV) layers followed by two fully connected (FC) layers. Similarly, the model for the HAR dataset includes one CONV layer followed by two FC layers.

To achieve efficient execution under constrained resources, we employ a pattern-based concentration method, which involves pattern pruning followed by a two-level fine-tuning process to recover lost accuracy, as discussed in Section II-C1. We select 3, 4, 6, and 15-entry patterns, resulting in approximately 10 total pattern sets. The intuition behind pattern selection is based on L1 normalization, where weights with magnitudes closer to zero are given higher priority for pruning. After pruning and multi-level fine-tuning, the most critical weights are concentrated within the pattern regions, achieving concentration ratios of 58.8%, 33.3%, and 44.44% for the MNIST dataset, as detailed in Table I. Similarly, a 24.3% concentration ratio is achieved in the FC layer of the HAR dataset. Lastly, for the GTSRB dataset, weight concentration reaches 44.44% in the CONV layers and 55.55% in the FC layers. The final compressed model sizes are 19K, 14K, and 33K parameters for MNIST, HAR, and GTSRB, respectively. The key idea behind weight concentration is that the compressed model is activated during low-energy adaptation, ensuring efficient execution.

Beyond the CONV and FC layers, the models also include pooling and batch normalization layers. However, since these layers occupy an insignificant amount of memory and introduce minimal computational latency, they are not explicitly discussed here. s shown in Figure 7 (a), each model undergoes a two-level fine-tuning process. Initially, the accuracy for MNIST, GTSRB, and HAR models is 97.2%, 86.3%, and 91%, respectively. After applying pattern-based sparsity, accuracy drops to 13.9%, 41.1%, and 15.3%, respectively. In the first fine-tuning stage, the concentrated weight regions are optimized, restoring accuracy to 93.4%, 81%, and 82.7%. In the second stage, the concentrated regions are frozen, and only the pruned weights undergo fine-tuning, to achieve similar to the initial accuracy.

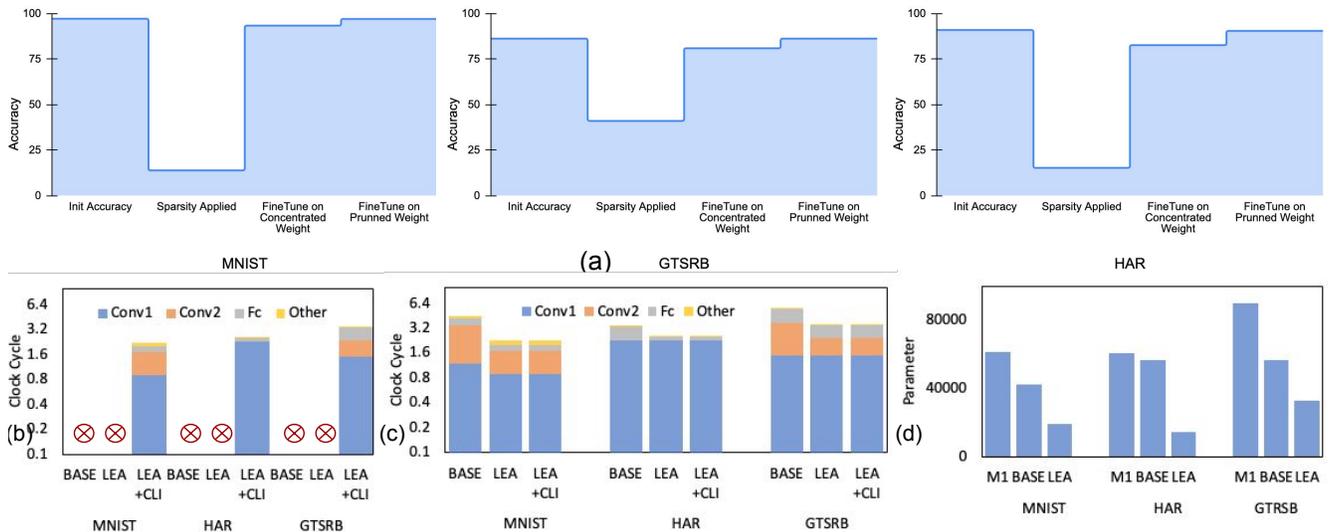

Fig. 7: (a) Accuracy preservation under several fine-tuning method (b) Inference under intermittent power supply (c) Inference under contiuous power spply (d) Memory consumption

*2) Inference under Continuous Power Supply:* Figure 7(c) illustrates the on-device inference latency for each dataset and model, measured in million clock cycles (Mcc) and displayed on a logarithmic scale for enhanced visibility. Here, BASE denotes the baseline model without low-energy adaptation (LEA) or checkpointless intermittent execution (CLI) mechanisms. LEA represents the model with concentrated weight patterns activated under low harvested power, while LEA+CLI combines both of the optimizations. For the evaluated datasets, the BASE model requires 450 Mcc, 340 Mcc, and 560 Mcc, respectively. In contrast, both the LEA and LEA+CLI configurations exhibit comparable latencies of 225 Mcc, 255 Mcc, and 350 Mcc, achieving an average 1.65× reduction in latency compared to the BASE model. This demonstrates the efficiency gains enabled by the proposed optimizations under energy-constrained conditions.

*3) Inference under Intermittent Power Supply:* To simulate diverse energy harvesting scenarios, we employed the SIGLENT SDG1032X function generator [21] coupled with a 100 μF buffer capacitor. As shown in Figure 7 (b) (logarithmic scale), the BASE and LEA models fail to complete execution under intermittent power conditions due to their lack of an efficient checkpointless intermittent execution framework.

In contrast, the LEA+CLI configuration successfully completes inference tasks. This is enabled by CLI, which provides loop indeices based intermittence support without checkpointing overhead. Therfore, the overall latency for each dataset and model remains nearly identical to the latency for continuous power supply. This demonstrates the practicality of combining energy-aware weight adaptation (LEA) with lightweight intermittent execution (CLI) for energy-harvesting devices.

*4) Memory Requirement:* Since the parameter size of a DNN is a critical bottleneck for memory requirements, we evaluated memory consumption by measuring the number of parameters. Figure 7(d) shows that the traditional setup (M1), which activates different models in response to low-energy conditions, requires 61K, 60.8K, and 90K parameters for the MNIST, HAR, and GTSRB datasets, respectively. In comparison, the BASE model, without energy adaptations, requires 42K, 56.8K, and 56.5K parameters, while our proposed LEA model, with concentrated weight patterns, significantly reduces the parameter count to 19K, 14K, and 33K for the same datasets. On average, the LEA model achieves 3.4× and 2.6× greater memory efficiency compared to M1 and BASE, respectively, demonstrating its effectiveness in minimizing memory overhead under resource-constrained conditions.

## IV. CONCLUSION

In this paper, we presented an energy-adaptive, checkpoint-free intermittent inference framework designed for low-power energy-harvesting (EH) devices. Our approach addresses three critical challenges: adapting deep neural network (DNN) inference to fluctuating energy conditions, ensuring computation consistency despite frequent power interruptions and lastly write-after-read (WAR) hazard during intermittent computation. We introduced a dynamic low-energy adaptation (LEA) mechanism that reduces computational complexity without the need for deploying additional models. By concentrating essential weights and utilizing pattern-based pruning with fine-tuning, LEA significantly improves memory efficiency and reduces inference latency. Additionally, our checkpoint-free intermittent inference method leverages loop index-based progress tracking, eliminating the overhead associated with traditional checkpoint-and-restore mechanisms. Lastly, through a consistency-aware idempotent execution strategy we solve the WAR hazard. Thus, we ensure consistent and reliable inference even under intermittent power conditions.